\documentstyle[aps,preprint,epsf]{revtex}

\begin{document}

\title{Voltage Distribution in Growing Conducting  Networks}

\author{Bosiljka Tadi\'c$^{1,*}$ and   Vyatcheslav Priezzhev$^{2,**}$
}

\address{$^1$Jo\v{z}ef Stefan Institute,
P.O. Box 3000, 1001 Ljubljana, Slovenia \\
$^2$Bogolubov Laboratory of Theoretical Physics,
Joint Institute of Nuclear Research, 141980 Dubna, Russia }


\maketitle
\begin{abstract}
\newline
We investigate by  random-walk simulations and a mean-field theory how
growth by biased addition of nodes affects flow of the current through
the emergent conducting graph, representing a  digital circuit.
 In the interior of a large network the
voltage varies  with the addition time $s<t$ of the node as  $V(s)\sim
\ln (s)/s^\theta$ when constant
current enters the network at last added node $t$ and leaves at the root
of the graph  which is grounded. The topological closeness of the conduction
path and shortest path through a node suggests that the  charged random walk
determines these global graph properties by
using only {\it local} search algorithms. The results
agree with mean-field theory on tree structures,
while the numerical method is applicable to graphs of any complexity.

\end{abstract}
\pacs{PACS numbers:89.75.Hc Networks and genealogical trees,
05.40.Fb   Random walks and Levy flights, 02.50.Cw   Probability theory,
89.20.+a   Industrial and technological research and development }


Networks, which are adequately represented by random graphs,
invade all sciences \cite{Strogatz,Reka,Dorogovtsev}.
 In a classical approach,  random graph
theory deals with linking in a static graph with a given number
of nodes \cite{book}.
Recently dynamically evolving networks came into focus \cite{Reka,Dorogovtsev}
representing connections in  complex dynamical systems, e.g., metabolic
or protein networks, and  realistic social and technological
networks, Internet and the Web, which are not static but evolve in time.
Details of the growth rules, in which new and/or preexisting nodes are
linked to the network, determine the graph topology that
emerges after long evolution time. In the  class of
scale-free networks preference attachment rules lead to power-law
degree distributions of incoming  and outgoing links
\cite{Reka,Dorogovtsev,BT}.

Conducting networks, such as electrical or electronic circuits are of
particular  importance for technology. A common technological network---digital
circuit---consists of logic gates, as nodes, and wires in a broad sense, as
links \cite{FJS}. Technology advance with integration and miniaturization
allows digital circuits to grow in size and complexity in order to
optimize their function and stability. It was shown recently \cite{FJS}
that electronic circuits exhibit a scale-free link structure up to a
cut-off size. So far, electrical properties of growing conducting
networks have not received much attention in the literature.

Here we adapt the random-walk dynamics and mean-field theory to study
for the first time how the growth of a conducting network interferes with
the current flow through the underlying {\it evolving graph}.
In particular, we study voltage
distribution per node when the unit external current  flows through the
network of conducting links, which are systematically  added in time  and
 attached  to the network with a preferential rule.
We grow a network and let  the constant current
flow into  last-added node and leave the graph  at
its  root (first node), which  is  kept at zero potential.
Compared to the static case, in the emergent graph
structure the time {\it when} a node was added to the graph determines
{\it how} it will be connected. By subsequent addition of nodes
both number of links grows and a new structure  emerges among
preexisting nodes, making the conduction path between
last-added node and root  fluctuate.
These features affect conduction on an evolving network
that we address in this work.
 To elucidate all aspects of the evolution,  we simulate suitable random
walks on networks of several sizes $N$, i.e., after  $t=N=2^k\times 500$
added nodes, $k=0, \cdots ,4$, and determine the universal
voltage curve that solely depends on time when the node was added to the graph.
For the graphs with tree structure we develop a mean-field theory
which qualitatively describes the numerical data.

Starting from  the  first node, the network grows by addition of one node
and one link per time step. The link is directed
to  one of preexisting  nodes $s$ with the probability
$p_{in}(s,t) = {{\alpha +q_{in}(s,t)}\over{(1+\alpha )t}}$,
where $\alpha \leq 1$ is a parameter and $q_{in}(s,t)$ is the number of
in-links of the recipient node $s$ at the moment $t$. Similarly, in the
general case an out-link at time $t$  occurs from new added node with
probability $g$,
whereas with $1-g$ it is a rewiring link from an earlier node, which is
selected with probability \cite{BT}
$p_{out}(s,t) = {{\alpha +q_{out}(s,t)}\over{(1+\alpha )t}}$.
By solving the corresponding rate equations
with the right boundary conditions $q_{in}(s,s= t)=0$, and
 $q_{out}(s,s=t)=g$,  we have
that the number of links per node increases in time as
\begin{equation}
q_\kappa (s,t) = A_\kappa \left[\left(t\over{s}\right)^{\gamma _\kappa}
-B_\kappa \right] \ .
\label{qst_ga}
\end{equation}
Here index $\kappa $ means 'in' or 'out' and the corresponding constants are
$A_{in}= \alpha$, $B_{in} = 1$, $A_{out} = \alpha+g$, $B_{out} =
\alpha/((\alpha +g)$, and the
exponents $\gamma _{in} = 1/(1+\alpha )$ and $\gamma _{out} =
(1-g)/(1+\alpha )$, respectively.
 Note that for $g<1 $ rewiring among the preexisting nodes occurs, which is
 the mechanism that leads to the hierarchical structure of out-links,
as demonstrated for instance in the model of the world-wide Web \cite{BT}.
In this case a number of closed cycles on the graph occurs. For $g=1$,
however, we are left with
tree structure of the graph  and Eq.\ (\ref{qst_ga}) for out-links
reduces to $q_{out}(s,t) = \alpha $ at  all nodes in the network.
In what follows we will mainly discuss the case $g=1$ and $\alpha =1$
where we have \cite{q_comment}
$q(s,t) = q_{in}(s,t)+ 1 = \left(t\over{s}\right)^{1/2}$.
We grow an ensemble of networks with these parameters on computer (an
example with first 39 nodes is shown in Fig.\ 1).
 The computed   average number of links $q(s,t)$ per node
after $t=N=8\times 10^3$ evolution steps is shown as top curve in Fig.\ 2,
which agrees well with the exact expression in Eq.\ (\ref{qst_ga}).
As the network  grows we fix the elements of the adjacency matrix $\hat{A}$,
so that after $N$ steps we have an $N\times N$ matrix with elements
$a_{xy} =1$  when a link $x\to y$ occurs, and zero otherwise.
Here we assume that these links are conducting in both directions.

An electrical network can be regarded as a graph in which the resistance
$R_{xy}$ is associated to the edge (link) between each pair of connected
 nodes $x\to y$. When two points (nodes) $a$  and $b$ of the graph are
connected to poles of a
 battery, the current and the voltage in the interior of the graph
are governed by the  Kirchhoff's laws. In particular, when
the potential difference occurs between points $x$ and $y$, the
current is given by the Ohm's law $i_{xy}=(V_x-V_y)C_{xy}$, where
$C_{xy}=1/R_{xy}$ is
the conductance of the respective link. By the Kirchhoff's current law
total current outflow from any point in the interior
is zero, $\sum _yi_{xy}=0$, we then find  for the voltage
\begin{equation}
V_x = \sum _y V_yC_{xy}/C_x  \ .
\label{Vx}
\end{equation}
where $C_x = \sum _yC_{xy}$ and the sum is over all nodes $y$  which are
connected to $x$.

The averaging property expressed by Eq.\ (\ref{Vx}) implies that the
voltage is a {\it harmonic} function on the interior points of the graph.
This makes the basis for the probabilistic interpretation of the
voltage \cite{book,scripta}. Namely, one can define another harmonic function,
e.g., by using the random walk \cite{BT_ARW} on the graph,
with the same boundary values.
The random walk determined by the electrical network is defined as an
(ergodic reversible) Markov chain with the transition probabilities $P_{xy}$
that are weighted with the conductances as $P_{xy}=C_{xy}/C_x$.
Then, when the constant voltage is applied to the graph such that
$V_a=1$ and $V_b=0$, the voltage in an interior point $x$ is determined
as the hitting probability $h_x$ that a walker staring at $x$ reaches
the point $a$ before reaching $b$.
In the scenario, which we also use in this work, when a  constant current
flows into the network at the point $a$ and leaves at $b$ the walk begins
at $a$ and is trapped when it reaches point $b$.
The harmonic function which is equivalent to
the voltage is then given by  \cite{book,scripta}
 $V_x=u_x/C_x$, where $u_x=\sum _yu_yP_{yx}$ is the expected
number of visits of the walker to point $x$ before it reaches $b$.
Consequently, the current
between interior points $x$ and $y$ is given by the {\it net}
 number of walks along the link between these two points.

In the network evolving for $t=N$ steps we apply the unit current flowing
into the network at the last-added node ($a\equiv t$) and leaving it at
the first-added node ($b\equiv1$). We assume that all
resistances are equal $R_{xy}=1$ and the walker moves both along out-links
and against in-links with equal probability. Therefore $C_x=q(x,t)$, the total
 number of links attached to node $x$ at time $t$. Hence  the voltage
at node $x$ is
\begin{equation}
V(x,t) = <u(x,t)/q(x,t)> \ ,
\label{voltage}
\end{equation}
where $u(x,t)$ is the number of visits at $x$ made by the walkers
starting at last node $t$  {\it before} they are trapped at node $1$.
The averaging in Eq.\ (\ref{voltage}) is over entire ensemble of walkers.
As mentioned above, we simulate random walks after $t=N=2^k\times 500$,
$k=0,1, \cdots , 4$ evolution steps.  We generate 400 different networks of
the same size $N$ and use 200 walkers at each
network realization, hence the voltage
in Eq.\ (\ref{voltage}) is determined using 80000 walkers. (Note that
the simulated voltage appears to be normalized by a constant $L\sim \ln{(N)}$
related to the average length of the conduction  path.)
The results for the network after $N=8000$ added nodes are given in Fig.\ 2.

The power-law dependence of the average degree  $q(s,N)$ on addition
time $s$ of a node (cf. Fig.\ 2 and  Eq.\ (\ref{qst_ga})) manifests the
basic property of the evolving networks with emergent scale-free
structure, where the most connected nodes are  those added to the network at
earlier stages of the growth. It appears that the average
number of visits per node also exhibits a power-law decay with the
time of addition $s$ with an exponent  $\theta _u \approx
0.58(3)$ (see Fig.\ 2). A sharp exponential cut-off  at
recently added nodes  $s\leq N$ suggests lack of links, that will appear
only in later stages as the network continues to grow. We find the same
finite-size effect on the voltage curve. However, the power-law
dependence here is modified by a logarithmic term. Apart from the last point
$s=N$ where (normalized) voltage is  $V(N,N)=1$, the
 approximate expression fits the data for $1\leq s\leq N-1$ as
(see Fig.\ 3)
\begin{equation}
V(s,t=N) = D(N)\ln{(s)}s^{-\theta }\exp{[-1.3(s/N)^4]} \ .
\label{VsN}
\end{equation}
Here $D(N)\sim 1/\ln{(N)}$ and
$\theta \approx  0.25\pm 0.04$ for the range of network sizes $N$ used
in this simulations. In Fig.\ 3 are shown separate fits for several
simulated network sizes $N$.
The finite-size (finite evolution time) effects can be adequately
dealt with by rescaling of the respective curves by
$f(s/N)=\exp{[1.3(s/N)^4]}/\ln{(N)}$. The master curve representing the
scaled voltage as function of $s$ is shown in Fig.\ 3 (top panel).

The curvature of the universal voltage curve at early nodes and a subsequent
decay with the addition time $s$ can be related to the growth process as
follows.
 Due to the preference linking often a direct link from the
high-voltage node $N$ is attracted by the group of nodes near the
root, thus `pumping' voltage to these nodes. (Note that
in the absence of loops voltage decays linearly  with the number of
junctions along the conduction path, while the actual position of the
conduction path  fluctuates with  node addition). On the other hand,
the increase of the voltage is compromised by highly probable linking of
the early nodes  to the root node $1$. For large evolution times
the cluster of nodes having a path
to  node $1$ grows faster compared to  clusters
linked to other junction points along the conduction path (see Fig.\ 1).
Hence  for large $s < N$ the probability increases for  a node  $s$ to
belong to the dominant cluster, which has zero voltage.

In the mean-field approach a network of $t$ nodes can be regarded as
consisting of $M\approx \ln{(t)}$ layers of nodes, where each layer is
defined by the
distance from the first node in the origin $d(x\to 1) =i=1,2, \cdots M$.
Here the distance $d(x\to 1)$ is defined by  number of links separating
a node $x$ and node $1$.
The conduction path from last added node to the origin cats through these
layer making one junction point $j_i$ at each layer. It is clear that a
node which belongs
to layer $i$ is linked to a node on preceding layer $i-1$. However,
while the network grows a node added at time $t$ may be attached to one
of the already existing layers. Hence,
the population of layers $R_i$ grows in time following precisely the above
bias attachment rule, which leads to the rate equation \cite{TP_longarticle}
\begin{equation}
{{dR_{i+1}}\over{dt}} =  {{(R_{i+1}/R_i +\alpha )R_i}\over{t(1+\alpha )}} \ ,
\label{rate_R}
\end{equation}
with the initial condition $R_0=1$.    The system (\ref{rate_R}) can
be solved recursively yielding
\begin{equation}
R_i(t) = (-\alpha)^i + t^\chi \sum _{\ell =0}^{i-1}{{K_{i-\ell}}\over{\ell !}}
\left({{\alpha \ln{(t)}}\over{1+\alpha }}\right)^\ell \ ,
\label{Ri}
\end{equation}
where $\chi =1/(1+\alpha) =1/2$ in the present case. The condition
that $\sum _{i=0}^mR_i(t) = t = e^m$ at current moment of time $t$,
 leads to the recursion relation between the coefficients. For instance,
for $\alpha=1$
we have $K_m=e ^{m/2} -e ^{-m/2}(1+(-1)^m)/2 -
\sum _{\ell =1}^{m-1}K_{m-\ell}\sum_{\kappa =0}^\ell (m/2)^\kappa /\kappa !$.

Now consider a subgraph $G_s$ at the moment $s< t$ of the graph $G_t$ grown
for $t$ steps. The number of nodes in the subgraph $G_s$,  $x \leq s$
that are on the
distance strictly larger than a given distance $i$,  $d(x\to 1)>i$, is
$n(d(x\to 1)>i) = s-\sum_{k=0}^iR_k(s)$. Then the probability that
the node added at the moment $s$ is among them reads
$Prob(d(s\to 1)>i)=1-\sum_{k=1}^iR_k(s)/s$.
In addition, if that node is on a path that hits the conduction path
 at a junction {\it above} layer $i$, then it has the
voltage $V_s>i$. Then the probability of voltage $ V_s>i$ is
\begin{equation}
Prob(V_s>i) =\left(1-\sum_{k=1}^{i}R_k(s)/s\right)/R_{i+1}(t) \ ,
\label{Vsgi}
\end{equation}
where, for short, $V_s\equiv V(s,t)$.
Notice that the conduction path is set by addition of the last node $t$
in the entire graph. Therefore
the probability $1/R_{i+1}(t)$ that the path from $s$ to $1$ does not miss
the junction point on  $(i+1)$th layer depends on the
population of that layer at the moment $t$. Combining the two
probabilities we have
\begin{equation}
Prob(V_s=i)  =Prob(V_s>i-1)-Prob(V_s>i) \ ,
 \label{Vs_mf}
\end{equation}
and the average voltage is given by  $V_s = \sum _{i=0}^M iProb(V_s=i) $.
Expanding the sum and using the corresponding expressions for $R_i(s)$
for $\alpha =1$ we
find  $V_s\approx c_0(1-1/s)  +c_1 s^{-1/2}\ln{(s)}
+ c_2s^{-1/2}(\ln{(s)})^2 + \cdots $. Here  $c_k$ depend on number of nodes
$t$. In particular, $c_0$ decreases for large $t$ and additional
higher-order terms  $\ln{(s)}^k$ appear. This series contributes to the
effectively reduced exponent of $s$ below $1/2$,
that justifies  the approximate fitting expression given in caption to
Fig.\ 3 (top).
More details  will be given elsewhere \cite{TP_longarticle}.

The simulated probability distribution of voltage obtained by the random-walk
statistics is shown in  Fig.\ 4. It can be fitted with a stretched-exponential
function. The probability distribution of survival time $t_w$ of the
random walk before trap, $P(t_w)$, and of the
frequency of visit $u$ of walkers to a given node, $P(u)$, appear
to have power-law dependences
with cut-offs, the latter resembling  closely the topology of shortest paths
on the graph \cite{Koreans}.

We have demonstrated how the electrical properties of an emergent graph
are shaped by the nature of linking processes governing
growth of the network. This establishes technologically relevant link
between, e.g., conduction of complex digital circuits and the way that
they are grown.
The main results are summarized in the dependence of the
voltage distribution per node inside the graph on the addition time of
the node to the network. Our mean-field theory qualitatively describes the
numerical data for the scale-free graphs with tree structure
 considered in this work.
On the level of the random walk with a trap, which is determined by the
electrical network, the observed voltage distribution
can be related to power-law dependences of the wandering time of the walk
and number of visits to a given node. As a side result we
have shown that the universal scaling exponent of the distribution of
visits of our ``charged'' random walk, which  uses local navigation rules,
coincide with the ones of the  distribution of shortest paths through
a node, for which costly global navigation is necessary. Thus,
close relationship between minimal path and conduction path on
an electrical network suggests potential use of  ``charged'' random walk to
determine global topological properties  by using local search algorithms
only. The numerical method is applicable to graphs of any link
complexity, for instance as given by Eq.\ (\ref{qst_ga}).

\acknowledgments
Work of B.T. was  supported by the Ministry
of Education, Science and Sports of the Republic of Slovenia.
Work of V.P. was supported by RFBR Grant
No. 99-01-00882 and SNSF Grant No.7SUPJ62295.


\narrowtext

\begin{figure}
\epsfxsize=76mm\epsffile[79 226 532 568]{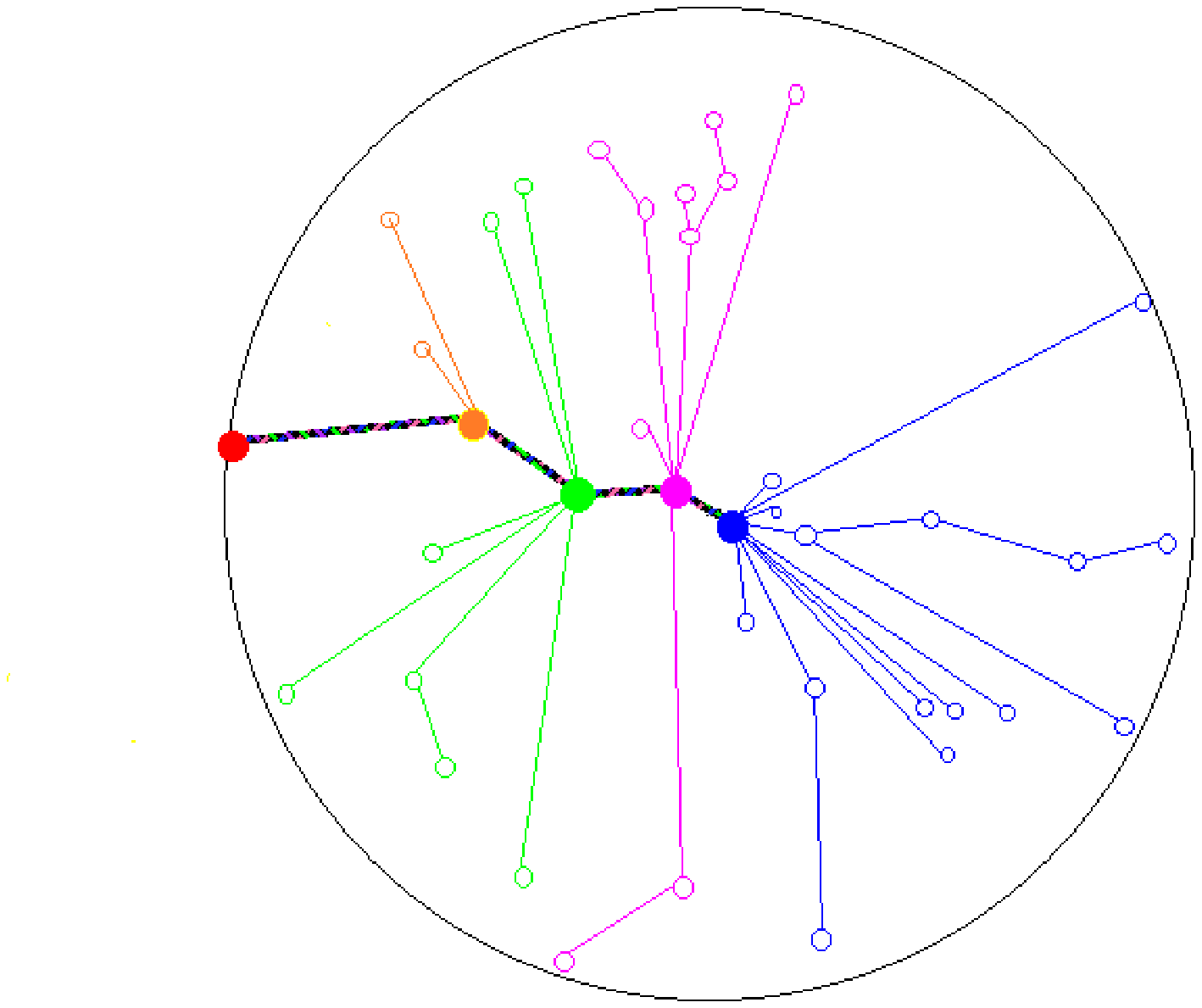}
\caption{\label{fig1}A network growing with the preference rule shown after
39 steps. Distance of a node (small circle) from the very first node in the
center illustrates the moment of its addition to the network. Current flows
along the conduction path (bold line) from the most recent node on the left
towards the center. The clusters of nodes meet the
conduction path at junction nodes which are shown by bullets.}
\end{figure}

\begin{figure}
\epsfxsize=76mm\epsffile[43 70 507 458]{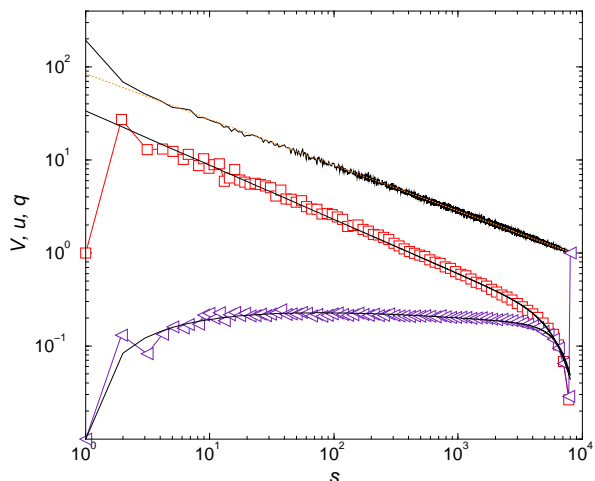}
\caption{\label{fig2}Average total degree $q(s,N)=1+q_{in}(s,N)$ per node
(top),  number of visits $u(s,N)$ (middle) and
voltage $V(s,N)$ (bottom) per node against time $s$ when the node was
added to the network with $N=8000$ added nodes.  Also shown are
 exact result $q(s, N)= (N/s)^{1/2}$ (dotted line) and fits
$u(s,N) = 33.6 s^{-0.58} \exp{[-1.4(s/N)^3] }$, and
$V(s,N) = 0.142\ln{(s)}s^{-0.23} \exp{[-1.3(s/N)^4]}$. Exact value
$V(s=1)=0$ was moved to a finite $10^{-2}$ to enable
presentation on the logarithmic scale. Data for $u(s,N)$
and $V(s,N)$ normalized to total number of walkers and log-binned. }
\end{figure}

\begin{figure}
\epsfxsize=78mm\epsffile[43 70 506 639]{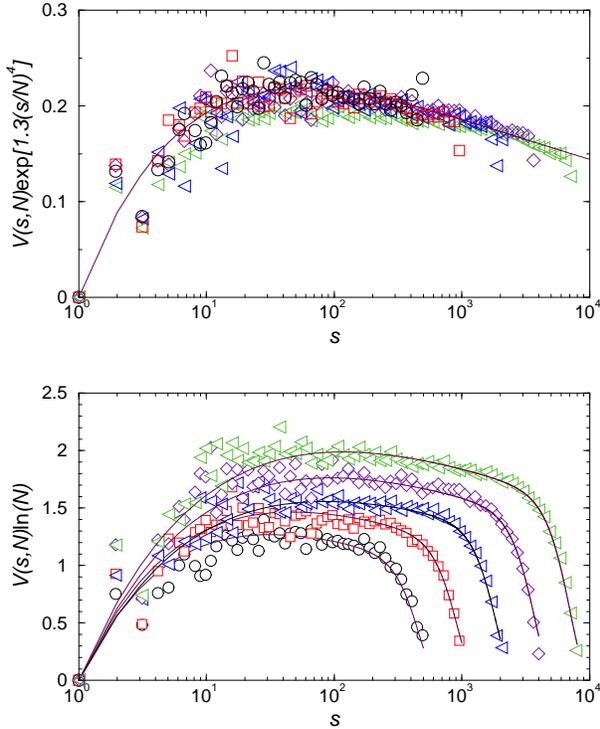}
\caption{\label{fig3}Lower panel: Average voltage per node
$V(s,N)\times \ln{(N)}$ vs. addition time $s <N$ obtained by
random walks on networks after $N=$ 500, 1000, 2000, 4000, and 8000 added
nodes (bottom to top).
 Full lines: respective fits according to Eq.\ (\ref{VsN}). Top panel:
Scaled voltage per node, normalized by $\overline{D(N)}$.
Full line: $y=0.02(1- 1/s) +0.135\ln{(s)}s^{-0.25}$. Data are log-binned. }
\end{figure}

\begin{figure}
\epsfxsize=76mm\epsffile[41 72 506 488]{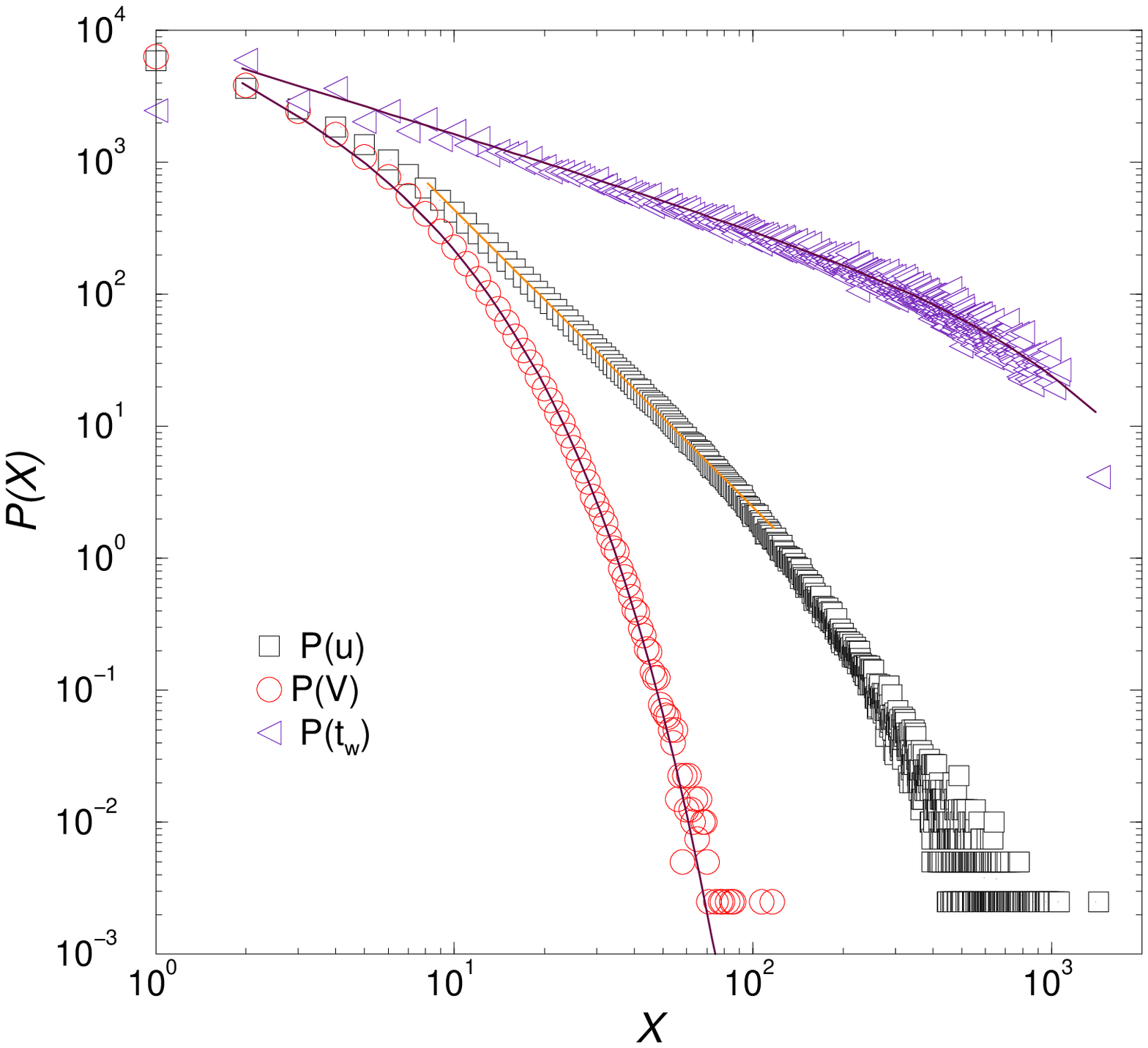}
\caption{\label{fig4} Probability distribution $P(X)$ of voltage
$X\equiv V$, and of number of visits $u$ and elapsed time before trap $t_w$
of random walkers on the network with $N=10^3$ nodes. Solid lines: fits
$P(V)= P_0V^{-0.8}\exp{(-V^{0.9}/3.8)}$,
and $P(t_w)= P_0t_w^{-0.75}\exp{(-t_w/N)}$, with $P_0=11\times 10^3$
and a power-law fit of the part of P(u) curve with slope 2.25$\pm$0.015.}
\end{figure}



\begin{references}
\bibitem{q_comment}In this case the degree distribution decays as
$\sim q^{-3}$, which was also found in electronic circuits in Ref.\ \cite{FJS}.
\end{references}
\end{document}